# Experimental observation of nanojets formed by heating the PbO-coated Pb clusters

Fengqi Song[1,a], Min Han[2], Minda Liu[2], Bang Chen[2], Jianguo Wan[1], Guanghou Wang[1,a]

1 National Laboratory of Solid State Microstructures and department of Physics, Nanjing University, 210093 Nanjing, P. R. China

2 National Laboratory of Solid State Microstructures and department of Material Science and Engineering, Nanjing University, 210093 Nanjing, P. R. China

We are reporting the first experimental observation of nanojets formed by heating PbO-coated Pb clusters, which has been predicted theoretically by M. Moseler and U Landman. During heating, the hot liquid is ejected through the broken orifice into a vacuum and forms a condensed trail in the shape of a tadpole, as shown in the TEM micrographs. The temperature-variable Raman spectra indicates that nanojet formation is closely related to heating temperature and thus essentially to the pressure in the coated clusters. The pressure inside the shell, which rises from the inner core's melting and its confined volume expansion. It then drops after the final explosion, dominating the whole nanojetting process.

---

[a]  To whom correspondence should be addressed
 Email: jackkiesong@vip.sina.com
   Fax: +86-25-83595535          Tel: +86-25-83595082

The formation of liquid jets and their breakup have been hot topics over the years; because of their potential application in ink-jet printing, spraying and coating, optoelectronic device manufacturing and microfluidics [1, 2]. Given the current trend of miniaturization, it is natural to consider the possibility of nanojet formation. It has been long believed that nano-sized jets could not be generated due to enhanced capillary force and viscosity in the nanoscale [1, 2, and 3]. Recently a molecular dynamic simulation has demonstrated that nanojets could be generated by ejecting propane through a very small nozzle with a solid gold wall [4]. By using a 22nm-diameter inlet-tube with a 6nm-diameter convergent nozzle, the inner propane under the constant pressure of 500Mpa was expulsed with a velocity of more than 200m/s and cooled down by evaporation with a trail formed.

Here are some key points for nanojets formation, such as a nanoscale-volume container confining the fluid under high pressure and a nozzle of nanometer size working for a sufficiently long time without becoming blocked. Physically, the complete implementation of these conditions is still a great challenge to technology. However, the development of nanoscience and nanotechnology has already shed light on these problems. Progress on the syntheses of nanoparticles and nanotubes have made it possible to form core-shell structures in spherical or columnar geometries. It was proposed that such composite structures could be used as reaction cells to investigate physical and chemical behaviors under extreme conditions in nanoscale [5]. When a coated cluster is sufficiently heated, a liquid core and intrinsic pressure are generated within the coated shell. This provides the possibility for experimental implementation of the predicted nanojets by thermally processing the core-shell structure.

In this report we will present experimental evidence that nanojets can be achieved by breaking PbO-coated Pb clusters via heating and discuss the dynamics of nanojet formation. When metal clusters are exposed to air, oxidation often occurs on their surfaces and then they become coated with thin oxide shells and metal cores. Generally the oxide shell is not uniform and easy to break. In the PbO-coated Pb clusters the inner Pb core will melt ahead of the outer PbO shell when it is heated and the internal pressure will keep increasing because of the liquid core's temperature-dependent volume expansion. Eventually the PbO shell is broken and the liquid jet will finally be formed. The process has been studied by Raman Scattering and TEM observations.

The Pb cluster beam was generated with a gas aggregation source and deposited on the surface of the TEM carbon film under ultra-high-vacuum (UHV) conditions. The cluster films were exposed to air for several days before UHV annealing. In situ TEM observation was carried out during the annealing process. When the film was heated up to 200ºC, the melted Pb within the PbO shell was found to burst out of the nanojets in the shape of tadpoles and can be observed on the TEM images [Fig 1]. The ejected materials form a condensed non-crystalline trail on the substrate surface. The sizes of the coated clusters are normally tens of nanometers and the lengths of the trails reach 240nm or even longer [Fig 1a, 1c]. Estimating from the trail length and considering evaporative cooling, we are convinced that a rather large amount of core material has been ejected. Indeed, even small clusters can be broken to form smaller trails if the temperature is high enough. Such smaller jets with a trail of 20nm are shown in Fig 1b. Moreover very small nanoparticles can be observed in the jet trail [Fig 1c] and they are a few nanometers in size. This will meet the prediction of the MD simulation [4]. TEM observation shows that the process of jet formation and condensation is finished in less than one second.

Raman scattering spectra were used to study the heating process of coated clusters. The measurements were carried out on a Raman spectrometer JY HR800 equipped with a micro-sensor with the radius of selected domain as small as several micrometers. The bottom curve of fig. 2a shows the accumulated Raman scattering spectrum of pure lead oxide powder (massicot) at 30ºC. The two peaks shown in the Raman spectra correspond to the $A_g$ and $E_u$ mode respectively, as assigned in [7]. The pure lead oxide powder was heated from 30ºC to 500ºC and the process was monitored by Raman scattering. No obvious change was observed for the positions and the FWHM of the Raman peaks during the heating process. However, this is not the case for the cluster films shown in Fig 2a. The upper three curves are the Raman spectra of the cluster film at different temperatures of 30 ºC, 220 ºC, and 390 ºC respectively. The spectra of the oxide shell can be obtained since the Pb core is Raman inactive. At 30ºC the Raman peaks of the cluster films are broadened in comparison with ones of pure lead oxide powder, which should be attributed to the interface effect in the nano-granular film [8]. Compared to the peaks at 30 ºC, the ones at 220 ºC obviously broaden. Furthermore, a series of temperature-variable Raman scattering spectra of the cluster film was accumulated from 0°C to 550°C. Fig 2b gives that the FWHM of the $A_g$ peak is plotted against the temperature. The peak width grows slowly from 6.3 cm$^{-1}$ at 0°C to 6.8cm$^{-1}$ at 100°C. Then the peak width rises abruptly and reaches a maximum of 11.8cm$^{-1}$ at 220°C, followed by a decrease down to the level of 9cm$^{-1}$ at 300°C, followed by a steady slight climbing. These measurements have been repeated and the sharp peak between 100~300°C is reproducible.

Peak width in the experimental Raman spectrum is composed of three parts: natural broadening, size-disorder broadening and pressure-dependent broadening. In general, the natural broadening remains stable while size-disorder increases steadily with the increase in temperature [8]. The temperature-dependent Raman peak-width curve in Fig. 2b obviously indicates that nanojet formation is pressure-related. Fig 2c schematically gives some hints of this process. The pressure keeps balance between the outer PbO shell and the expanding inner core by generating intrinsic force, leading to the broadening of the Raman peaks [9]. Therefore the sharp peak of Fig 2b curve reflects change in the internal pressure of the coated clusters. Pure Pb has a much lower melting point than PbO and the melting point of Pb clusters is even lower, estimated to be below 200°C [5]. During the heating process the inner Pb core will melt first at a rather low temperature and generate a large amount of pressure because volume expansion is subject to the confinement of the PbO Shell. This pressure is estimated to be as large as several hundred MPa [10]. In addition the liquid core will continue to expand and the internal pressure will rise accordingly with the increase of the heating temperature until the oxidized shell is broken and the pressure is released. Thus the sharp increase in the Fig 2b curve can be attributed to the melting and intermediate expansion of the inner Pb core whereas the following decrease may be due to the final break of the coated PbO shell. Therefore it is the temperature-dependent evolution of internal pressure in the coated cluster that drives the final nanojet formation.

We have also performed the UHV annealing processes of the films at different temperatures and investigated them by TEM. Fig 3a, b, and c show the TEM images of cluster films annealed at 150ºC, 200ºC and 250ºC for 20 minutes respectively. We can see that the morphology remains basically unchanged except for some conglomeration after annealing at 150ºC [Fig 3a]. However, the coated clusters explode and trails appear after annealing at 200ºC [Fig 3b]. Smaller clusters are found to break after annealing at 250ºC [Fig. 3c]. In fact clusters of larger size with the same-thick coated layer will break more easily [10].   Therefore the maximum internal pressure can be obtained by measuring the size of the smallest broken clusters. We have searched the TEM images for the smallest exploded clusters and their diameters are plotted versus the annealing temperatures in Fig 3d. With the increase of the temperature, the minimum size of exploded clusters decreases from 56nm at 180 ºC to 12nm at 250 ºC, which further confirms the internal pressure change in the core of the coated cluster mentioned above.

From the above observations the following points can be obtained. When the coated clusters are heated, the inner Pb core may begin to melt at first at about 160ºC [8] and then the internal pressure increases quickly [Fig 2b]. As the temperature continues to increase, the internal pressure increases further due to the confinement of continuous expansion of its liquid core. The larger clusters start to explode at about 200ºC (Fig 3b) with nanojet formation and smaller ones follow (Fig 3c). At 250ºC, clusters with the size of 12nm are broken. The internal pressure is finally released at about 250ºC when nearly all the clusters are broken (The sharp decrease in Fig 2b). Both the TEM observation and the Raman measurements agree well with our proposed jetting mechanism. Therefore, we hold that it is the temperature-dependent core pressure that drives the whole process and finally forms the jet in such systems. Secondly, the process of nanojet formation is more or less similar to the one described by theoretical prediction except for the permanent pressure used [4]. To estimate from the parameter of the bulk Pb at 25ºC [10], the pressure of 1000MPa will compress the bulk Pb by 2%. This means that the confinement of the inner core volume expansion will generate pressure in the order of $10^8$ Pa, the same order given by the simulation [4]. In addition the length of the jet trails are several tens of nanometers and small particles are formed in the trail. All these results suggest that this experimental nanojet formation share similar dynamic processes with the theoretical simulation [4].

In conclusion we have demonstrated that nanojets can be produced by heating coated clusters and the formation of nanojets is not only related to cluster size, but also closely related to heating temperature, which is in essence related to the temperature-dependent internal pressure of the coated clusters. Therefore heating the coated clusters to a low melting point is an effective method of producing nanojets and provides for the possibility of future nanodevice manufacturing. However the preparation of jet units of uniform configuration and regular pattern remains a great challenge and needs to be studied further.

Acknowledgements

This work was financially supported by the National Natural Science Foundation of China (Grant No. 90206033, 10274031, 10021001), the Foundation for University Key Teacher by the Ministry of Education of China (Grant No. GG-430-10284-1043), as well as the Analysis and Measurement Foundation of Nanjing University. We will also express our thanks to Dr X. Zhang and Prof G. Cheng for thoughtful discussions and to Brian Yarvin for improving English.

as reported in *J.Chem Soc Dalton* 1096 (1977).

8. M. J. Konstantinovac, S. Bersier, X. Wang, M. Hayne, P Lievens et al *Physical Review B* 66 161311(R) (2002). We have measured the ratio of the Stokes peak and anti-stokes peak and calculated the temperature $T$ by $I_S / I_{AS} = \exp(\eta\omega / kT)$, when $I_S$ and $I_{AS}$ mean the intensity of Stokes and anti-Stokes peak respectively. We have compared the calculated temperature with the one given by the thermocouple in the Raman spectrometer. The temperature modification from the 15mW-laser heating is determined to be about 40ºC. We can also judge the melting point of the Pb clusters by the sudden increase in pressure after correction from laser heating effect. The increase point is at about 120ºC and the correction is 40ºC. We can get that the melting point of the clusters in the selected domain is at about 160ºC.

9. As shown in Fig 2c, the tangent stress can be generated in order to create balance with the inner expansion. The stress will induce the Raman peaks' broadening. http://www.msm.cam.ac.uk/mmc/research/diamondfilm/diamond/diamond_results.html. by Debdulal Roy and T. W. Clyne

10. From the density of bulk lead, we have estimated that the melting of the Pb core will generate the volume expansion of 106%. At 25ºC the pressure of 1000Mpa will compress the bulk Pb by 2% i.e. the confinement of the inner core's volume expansion will generate the pressure with the order of $10^8$pa, the same order as in the theoretical simulation. Furthermore, the pressure will continue to increase because the liquid core will continue to expand with the increase of the heating temperature (For the free bulk liquid lead the temperature increase of 100ºC will generate 0.3% of the volume increase). Given the same thickness, the oxide shell will be broken from the same expansion. It is evident that with the same pressure, the larger cluster will generate a larger expansion. Therefore the size of the smallest broken cluster will present the pressure reached at the temperature.

Figure Captions

Fig. 1

TEM images of formed nanojets in heated cluster films.

Fig.2

A. Accumulated Raman spectrum of cluster film under different temperatures. The bottom one is the Raman spectrum of pure lead oxide powder. The other three curves are the spectra of cluster films at temperatures of 30 ºC, 220 ºC, and 390 ºC respectively. The intense peak of about 140cm$^{-1}$ corresponds to the $A_g$ mode [7].

B. FWHM of $A_g$ Raman peak versus annealing temperature.

C. A schematic of proposed force balance when the coated cluster is heated.

Fig. 3

A, B, C are TEM images of cluster films after 20-minutes UHV annealing at 150°C, 200°C and 250°C respectively. For a certain annealing temperature, frames of TEM images have been searched for the smallest exploded cluster. Their diameters are plotted against the annealing temperature in the inset D. It happened that the points form a straight line.

Fig.1 Song Fengqi et al

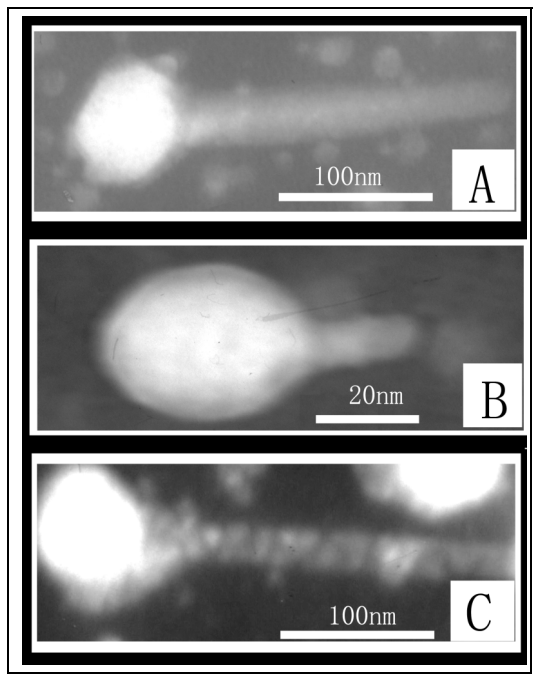

Fig.2 Song Fengqi et al

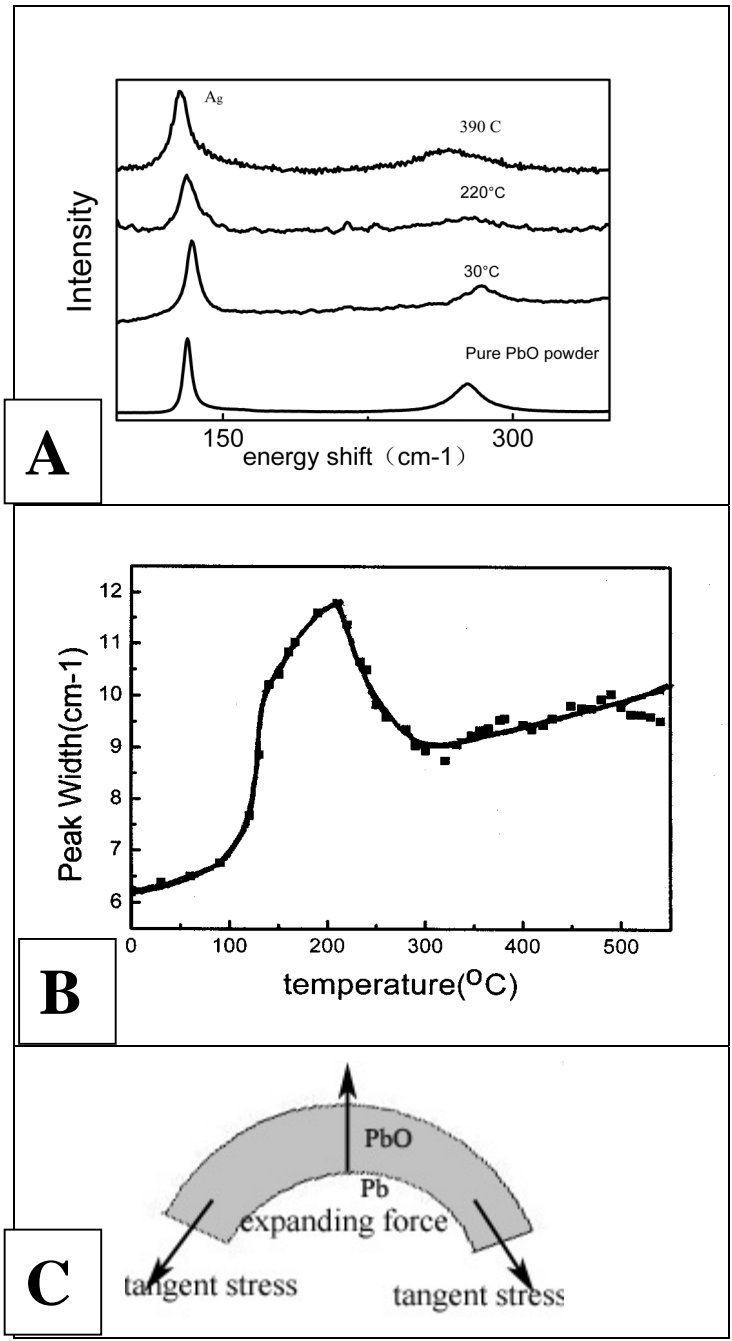

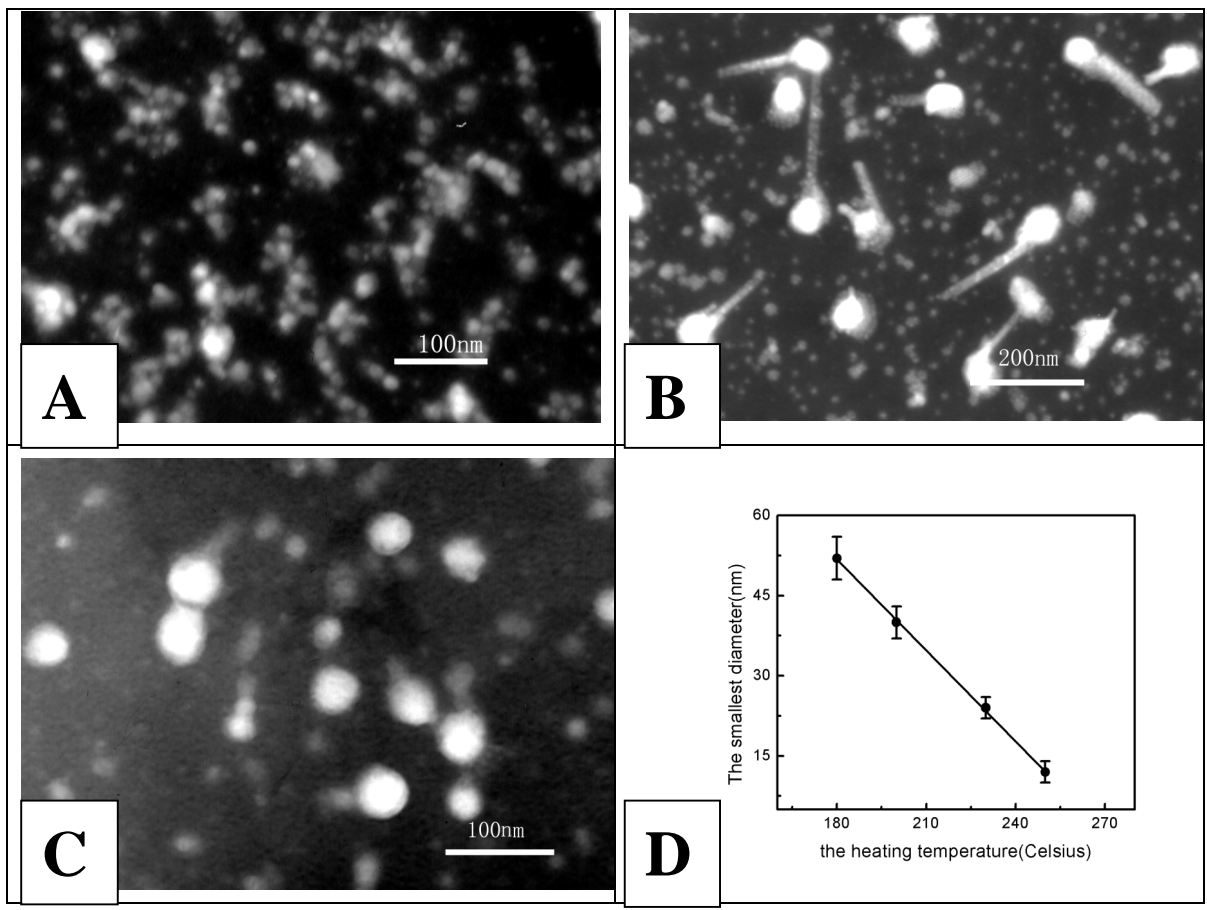

Fig.3 Song Fengqi et al